\documentclass[a4paper]{jpconf}
\usepackage{amssymb}
\usepackage{graphicx}
\usepackage{url}

\newcommand{\Kfo}{K_{\mathrm{fo}}}

\begin{document}
\title{Dynamical freeze-out in event-by-event hydrodynamics}

\author{Hannu Holopainen$^1$ and Pasi Huovinen$^2$}

\address{$^1$ Frankfurt Institute for Advanced Studies, Ruth-Moufang-Str.~1,
D-60438 Frankfurt am Main, Germany,\\
$^2$ Institut f\"ur Theoretische Physik, Johann Wolfgang
Goethe-Universit\"at, Max-von-Laue-Str.~1, 60438 Frankfurt am Main,
Germany}

\ead{holopainen@fias.uni-frankfurt.de}

\begin{abstract}
  In hydrodynamical modeling of the ultrarelativistic heavy-ion
  collisions the freeze-out is typically performed at a constant
  temperature or density. In this work we apply a dynamical freeze-out
  criterion, which compares the hydrodynamical expansion rate with the
  pion scattering rate. Recently many calculations have been done
  using event-by-event hydrodynamics where the initial density profile
  fluctuates from event to event. In these event-by-event calculations
  the expansion rate fluctuates strongly as well, and thus it is
  interesting to check how the dynamical freeze-out changes hadron
  distributions with respect to the constant temperature
  freeze-out. We present hadron spectra and elliptic flow calculated
  using (2+1)-dimensional ideal hydrodynamics, and show the
  differences between constant temperature and dynamical freeze-out
  criteria. We find that the differences caused by different
  freeze-out criteria are small in all studied cases.
\end{abstract}

\section{Introduction}

In hydrodynamical modeling of heavy-ion collisions the freeze-out is
typically assumed to take place instantaneously on a very thin
hypersurface of constant temperature or energy density, and the fluid
is converted to particles using the Cooper-Frye formula. However, from
microscopic point of view the system should freeze-out either when the
mean free path of particles is larger than the system size, or when
the expansion rate exceeds the scattering rate of
particles~\cite{Bondorf:1978kz}. Neither of these criterions is
directly proportional to temperature nor density. The dynamical
freeze-out criterion which directly compares scattering rate with the
hydrodynamical expansion rate was applied to hydrodynamical modeling
already some time ago~\cite{Hung:1997du}, but it has not been widely
used.
 
Earlier studies with smooth optical Glauber initial profiles have
shown that while the freeze-out surface from the dynamical condition
differ significantly from the constant temperature surface, the effect
on observable particle spectra is small~\cite{Eskola:2007zc}. However,
no studies of the effect on elliptic flow have been done yet. As well,
event-by-event hydrodynamical calculations have become popular
recently, and in such a case the flow develops more violently than
when averaged initial state is used. Thus it is not obvious whether
the difference between the two freeze-out conditions is small also
when the initial density fluctuates event-by-event.

The problem of freeze-out is not fully resolved in the hybrid models
either. In these models the hydrodynamical evolution is stopped while
the hadrons are still interacting, the fluid is converted to
individual hadrons, and the subsequent evolution of hadrons is
described using a hadron cascade~\cite{Petersen:2008dd}. The switch
from fluid to cascade is usually done on a constant temperature
hypersurface, but the final results depend on the temperature where
the switch is done~\cite{Song:2010aq,Hirano:2012kj,Huovinen:2012is}.
It is possible
that switching on a constant Knudsen number instead would lead to
results which are less sensitive to the actual value of the switching
parameter.

In this proceedings we compare the dynamical freeze-out criterion with
the constant temperature one. We start by introducing our
hydrodynamical model and the dynamical freeze-out criterion. We
compare the properties of the freeze-out surfaces obtained using
different freeze-out criteria and averaged initial state, and
calculate the particle $p_T$-spectra and elliptic flow in these
cases. Finally we apply the two different criteria to an event-by-event
calculation, study the surfaces, and calculate the particle
$p_T$-spectra and elliptic flow.

\section{Hydrodynamical model}

Here we use a slightly modified version of the ideal event-by-event
hydrodynamics framework presented in \cite{Holopainen:2010gz}. We
solve the ideal hydrodynamical equations
\begin{equation}
  \partial_\mu T^{\mu\nu} = 0 \qquad \partial_\mu N^\mu = 0,
\end{equation}
where $T^{\mu\nu}$ is the ideal energy-momentum tensor and $N^\mu$ is
the net-baryon number current. We assume longitudinal boost invariance
and thus the numerical problem reduces to (2+1)-dimensions. In order
to close the set of equations we need an equation of state (EoS) and
our choice here is s95p-v1 from Ref.~\cite{Huovinen:2009yb}.
 
Particle emission from the freeze-out surface is calculated using
conventional Cooper-Frye formula
\begin{equation}
  \frac{dN}{d^2p_Tdy} = \int_{\Sigma} f(x,p) p^\mu d\Sigma_\mu,
\end{equation}
where we integrate over the freeze-out surface $\Sigma$, $f(x,p)$ is
particle distribution and $p^\mu$ is the four momentum of the emitted
particle. The freeze-out surface elements $d\Sigma_\mu$ are found with
\texttt{CORNELIUS++}\footnote{\texttt{CORNELIUS} surface finding
  subroutine is available at the OSCAR code repository,
  \url{https://karman.physics.purdue.edu/OSCAR/}.} using the algorithm
described in \cite{Huovinen:2012is}.

After the thermal particle spectra are calculated using Cooper-Frye,
we sample particles using these spectra as a probability
distributions. Sampling is done as described in~\cite{Holopainen:2010gz}. 
Thus the number of each particle species is fixed, but the total
energy of particles in each event can fluctuate. Then the strong and
electromagnetic decays are done one particle at a time. In
\cite{Holopainen:2010gz} decays were done using \texttt{PYTHIA}, which
meant that all hadron resonances included in the EoS could not be
used. Here we use a separate decay program, which makes it possible to
include all hadrons present in the EoS.

In this contribution we consider Au+Au collisions at Relativistic
Heavy Ion Collider (RHIC) with $\sqrt{s_{NN}} = 200$~GeV. Smooth
initial conditions for the entropy density are obtained from the
optical Glauber model using a mixture of binary (75\%) and wounded
nucleon (25\%) profiles. For the fluctuating initial conditions a
Monte Carlo Glauber model with same mixture is used. Monte Carlo
Glauber model gives only the positions of the wounded nucleons and
binary collisions and we need to distribute entropy around these
positions before we can initialize hydrodynamical evolution. Our
choice is a 2-dimensional Gaussian:
\begin{equation}
    s(x,y) = \textrm{const.} \sum_{\textrm{wn,bc}} \frac{1}{2\pi \sigma^2} 
             \exp \Big[-\frac{ (x-x_i)^2 + (y - y_i)^2 }{2\sigma^2} \Big],
\end{equation}
where $\sigma$ is a free parameter controlling the width of the
Gaussian. Typical values for this fluctuation size parameter is order
of $0.5$~fm, our choices here are $0.4$~fm and $0.8$~fm. For
net-baryon density a pure wounded nucleon profile is used in both
cases. We choose the initial time to be $\tau_0 = 0.6$~fm.

Because there are differences between the two Glauber models and the
mixtures are not defined exactly in the same way, we cannot compare
the results obtained with smooth and fluctuating initial
conditions. Here we concentrate on the effects coming from the
freeze-out condition and comparisons between smooth and fluctuating
initial conditions with constant temperature freeze-out can be found
\emph{e.g.}\ from Ref.~\cite{Holopainen:2010gz}.

\section{Dynamical freeze-out condition}

For the system to maintain kinetic equilibrium, the scattering rate
should be much larger than the expansion rate. In other words
\begin{equation}
  K = \frac{\theta}{\Gamma} \ll 1,
\end{equation}
where $\Gamma$ is the scattering rate and $\theta$ is the
hydrodynamical expansion rate. As a ratio of (an inverse of) a
macroscopic length scale to (an inverse of) a microscopic length
scale, the ratio $K$ can be identified as a Knudsen number --- which
must be larger than one for hydrodynamics to be valid as well. Since
we are interested in freeze-out, the region of last rescattering, we
replace the vague requirement $K \ll 1$ by assuming that the system is
reasonably close to equilibrium until $K = \Kfo =1$, and test the
sensitivity of the spectra to the exact value of $\Kfo$ by using
values $\Kfo=0.5-2$.\footnote{In Ref.~\cite{Hung:1997du} it was argued
  that $\Kfo=2$ would give the region of the last rescattering.} Our
initial assumption $\Kfo = 1$ is a posteriori justified by a good
reproduction of the measured particle spectra. When we compare to
results calculated using freeze-out at constant temperature, we use
$T_\mathrm{fo}=140$ MeV for the same reason.

The expansion rate of a system is given by the hydrodynamical
model. For a boost-invariant system it is evaluated as
\begin{equation}
  \theta = \partial_{\mu} u^\mu = \partial_\tau u^\tau + \partial_x u^x 
           + \partial_y u^y +  u^\tau/\tau.
\end{equation}
We evaluate the scattering rate of pions in kinetically (but not
necessarily chemically) equilibrated hadron-resonance gas as
\begin{equation}
    \Gamma = \frac{1}{n_\pi(T,\mu_\pi)} \sum_i \int d^3 p_\pi d^3 p_i
      f_\pi(T,\mu_\pi) f_i(T,\mu_i) \frac{\sqrt{(s-s_a)(s-s_b)}}{2 E_\pi E_i} 
      \sigma_{\pi i} (s),
\end{equation}
where $n_\pi$ is the density of pions, $f_i(T,\mu_i)$ is the thermal
distribution function, $\sqrt{(s-s_a)(s-s_b)})/(2 E_\pi E_i)$ is the
relative velocity, $\sigma_{\pi i}$ is the cross section for
scattering of pion with particle $i$, and the summation over $i$ runs
over all particle species included in the EoS. Cross sections are
evaluated as in the UrQMD model \cite{Bass:1998ca}, \emph{i.e.}\ the main
contribution comes from resonance formation which is evaluated using
the relativistic Breit-Wigner formula. Details of this calculation
will be published in a later paper.

Scattering rates are different for different particle species, and
thus one could argue that freeze-out should take place when the
scattering rate of a particular species is equal to the expansion
rate, which would mean freeze-out at different times for different
particle species. However, to be consistent, this approach would
require us to modify fluid dynamics in such a way that when
\emph{e.g.}\ pions are decoupled, they are removed from the fluid. For
the subsequent evolution of the remaining fluid we should then include
the interactions with the non-equilibrated pion cloud, which is
clearly beyond the capabilities of fluid dynamics. Thus we make the
simplifying approximation that all the particle species decouple at
the same time, and since the pions are the most abundant particles, we
base our freeze-out description on the scattering rate of pions.

\section{Results with smooth initial conditions}

\begin{figure}[h]
\includegraphics[width=18pc]{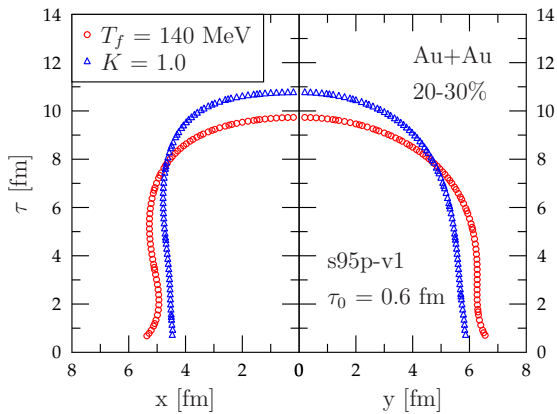}\hspace{2pc}%
\begin{minipage}[b]{14pc}
 \caption{\label{fig: surface comp} Freeze-out surface along positive
   $x$- and $y$-axis in 20-30\% central Au+Au collisions at
   $\sqrt{s_{NN}}=200$~GeV. Surfaces are shown with both dynamical and
   constant temperature criterion.}
\end{minipage}
\end{figure}

We start by studying the freeze-out systematics with smooth initial
conditions. First in Fig.~\ref{fig: surface comp} we compare the
freeze-out surfaces with both the constant temperature and the
dynamical freeze-out criteria. We can see that with the dynamical
condition the edges of the system decouple earlier and the center of
the system lives little longer. This is because near the edges the
pressure gradients are large and thus the expansion rate is large. In
the center of the system the expansion is not so rapid and thus the
center can maintain the kinetical equilibrium at lower temperatures,
\emph{i.e.}\ at smaller scattering rates.

\begin{figure}[h]
\begin{minipage}{16pc}
\includegraphics[width=16pc]{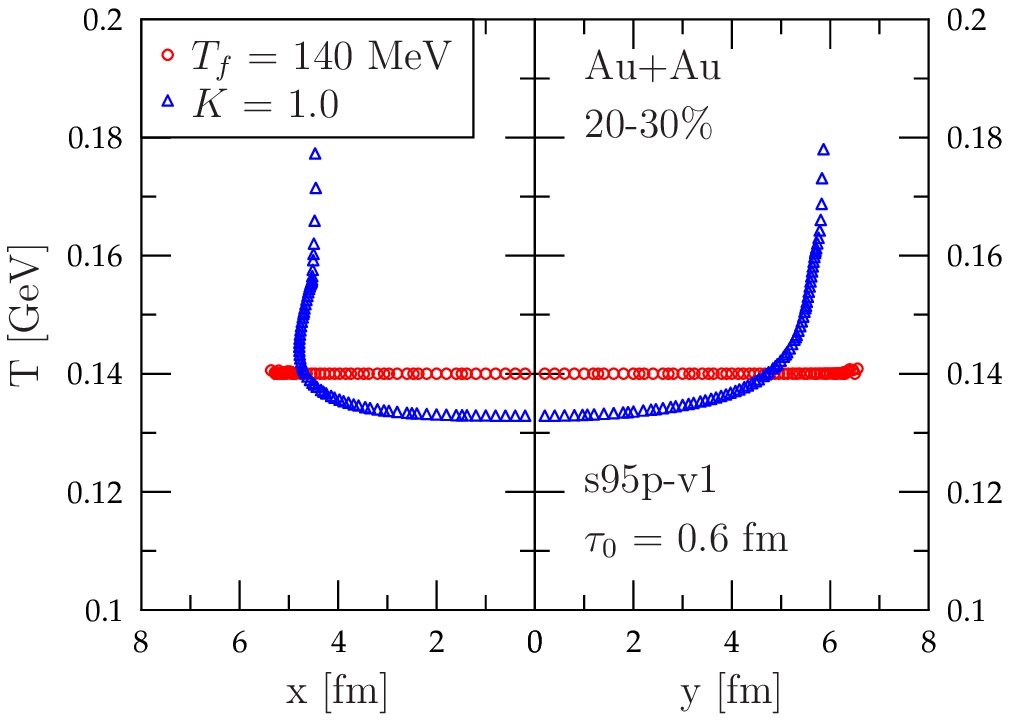}
\caption{\label{fig: surface temperature} Temperature on the
  freeze-out surface along positive $x$- and $y$-axis in 20-30\%
  central Au+Au collisions at $\sqrt{s_{NN}}=200$~GeV. Surfaces are
  shown with both dynamical and constant temperature criterion.}
\end{minipage}\hspace{2pc}%
\begin{minipage}{16pc}
\includegraphics[width=16pc]{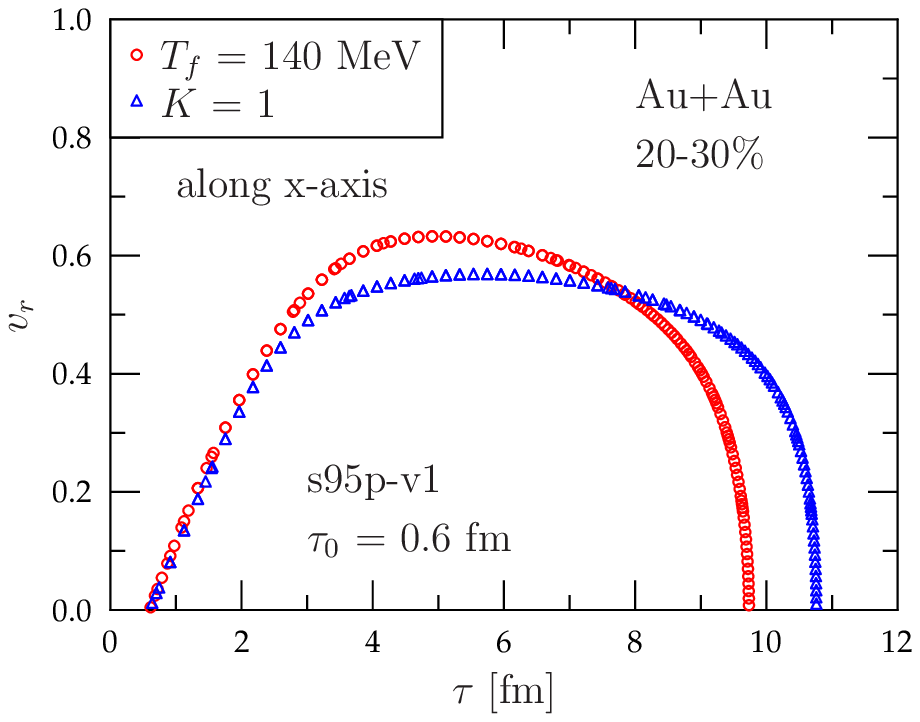}
\caption{\label{fig: surface velocity} Radial velocity on the
  freeze-out surface along positive $x$- and $y$-axis in 20-30\%
  central Au+Au collisions at $\sqrt{s_{NN}}=200$~GeV. Surfaces are
  shown with both dynamical and constant temperature criteria.}
\end{minipage} 
\end{figure}

In Fig.~\ref{fig: surface temperature} we plot the temperature on the
freeze-out surfaces with both conditions. Naturally with the constant
temperature criterion the curve is constant, but with the dynamical
freeze-out the temperature can vary a lot. As already explained above,
the edges decouple earlier with the dynamical conditions and thus the
edges are hotter than in the constant temperature case. Accordingly,
the center the system is little bit cooler with the dynamical
freeze-out.

The radial velocity on the surface along the $x$-axis as a function of
time is plotted in Fig.~\ref{fig: surface velocity}. Since with the
dynamical condition the edges decouple earlier, the largest flow
velocities are cut away as we can see in the figure. For the
high-$p_T$ ($\gtrsim 1$~GeV) emission the part of the surface where
the flow velocity is largest is most important and one could expect
that the two different freeze-out conditions could give different
results. However, the effect of smaller transverse flow is countered
by the larger temperature and thus it is difficult to say what happens
based on these figures alone.

\begin{figure}[h]
\begin{minipage}{16pc}
\includegraphics[width=16pc]{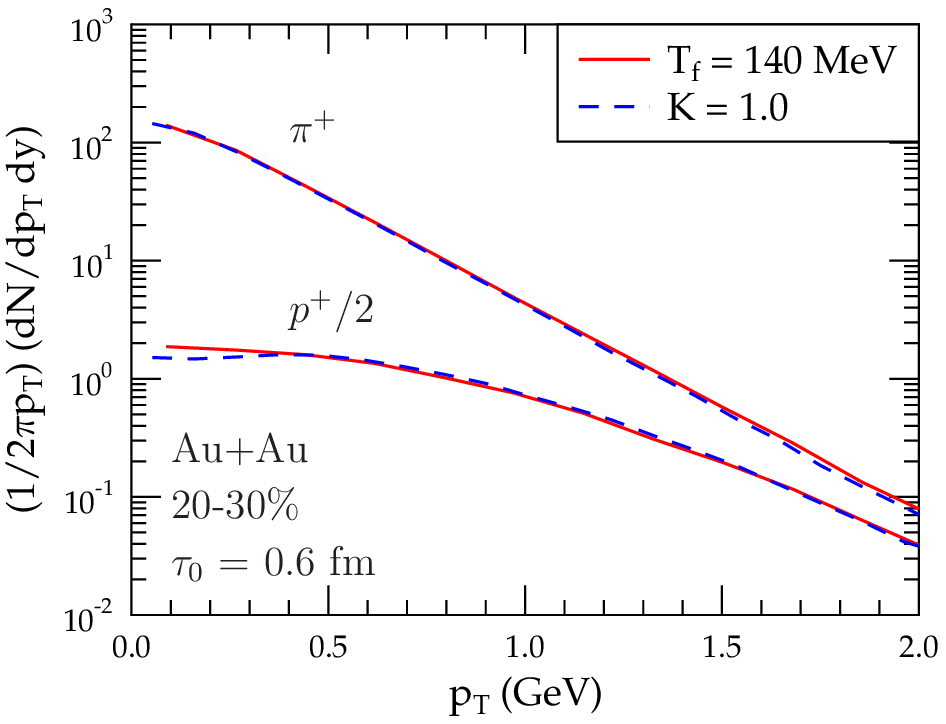}
\caption{\label{fig: spectra comp} Spectra of positively charged pions
  and protons in 20-30\% central Au+Au collisions at
  $\sqrt{s_{NN}}=200$~GeV. Results are shown with both dynamical and
  constant temperature criterion.}
\end{minipage}\hspace{2pc}%
\begin{minipage}{16pc}
\includegraphics[width=16pc]{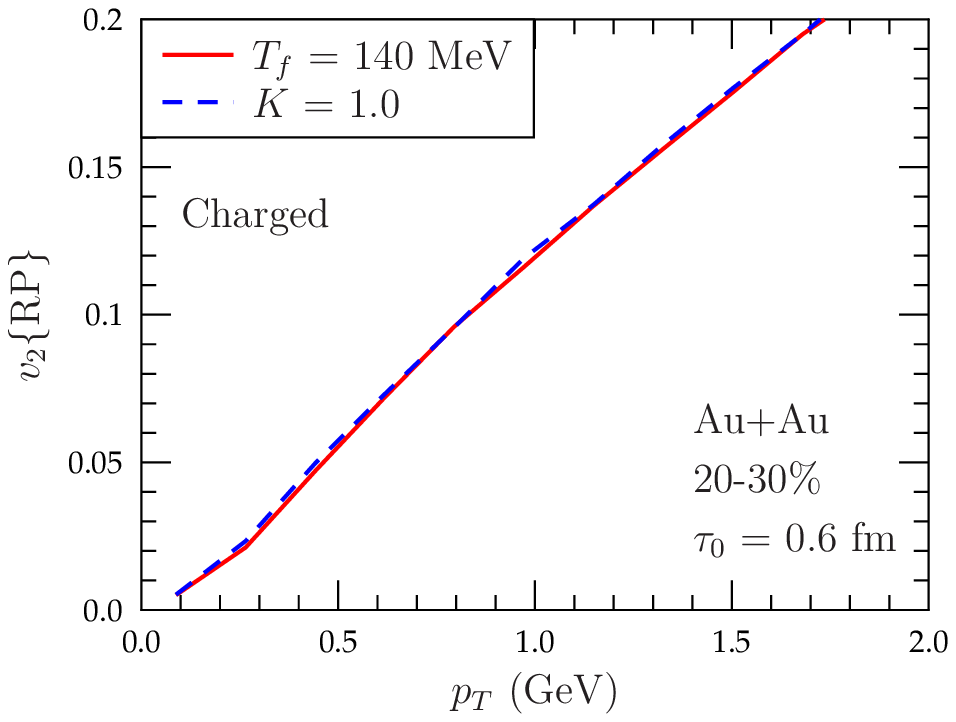}
\caption{\label{fig: v2 comp} Elliptic flow of charged particles
  calculated with event plane method in 20-30\% central Au+Au
  collisions at $\sqrt{s_{NN}}=200$~GeV. Results are shown with both
  dynamical and constant temperature criterion.}
\end{minipage}
\end{figure}

It is interesting to see how the freeze-out surface changes when we
change the freeze-out condition, but the experiments cannot measure
the properties of the freeze-out surface itself, but rather the
particle spectra and its anisotropies. In Fig.~\ref{fig: spectra comp}
we have plotted the spectra of positively charged pions and
protons. We see that results from different freeze-out conditions are
basically on top of each other. This indicates that the increased
temperature at the edges cancels the effect of reducing the maximum
flow velocity.

We also plot the elliptic flow of charged hadrons in 
Fig.~\ref{fig: v2 comp} with respect to the reaction plane (defined by
impact parameter and beam axis). Also here the freeze-out criterion
does not play a role and the result is the same with both
conditions. From the particle spectra and elliptic flow we can
conclude that the constant temperature freeze-out gives the same
answers as dynamical freeze-out despite the fact that the surface
itself shows some sensitivity to the criterion.

\begin{figure}[h]
\begin{minipage}{16pc}
\includegraphics[width=16pc]{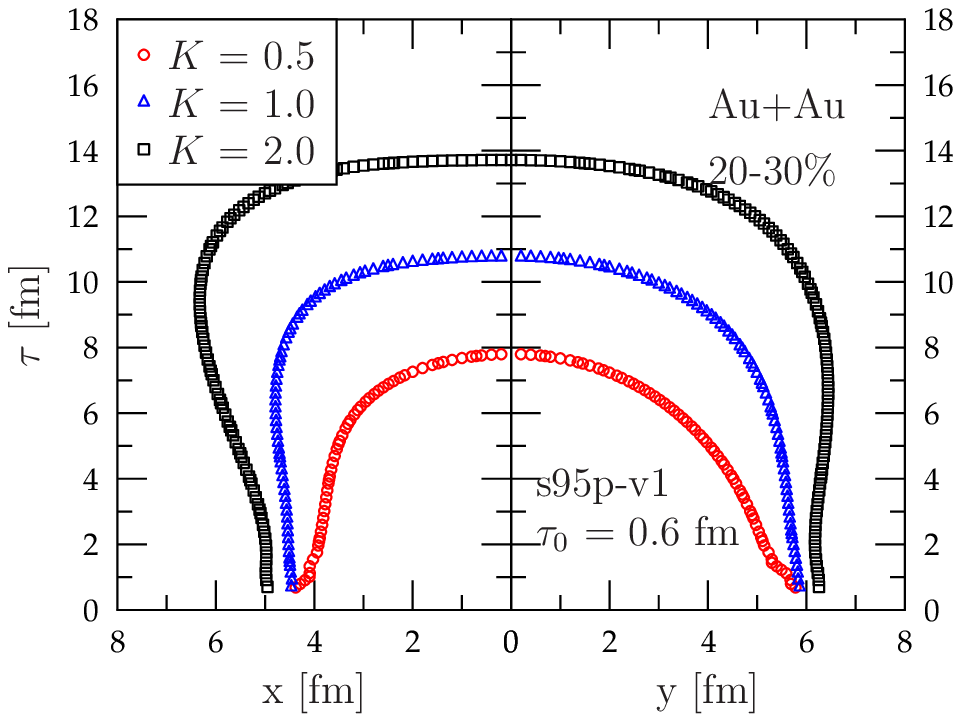}
\caption{\label{fig: chance K surface} Freeze-out surface along
  positive $x$- and $y$-axis in 20-30\% central Au+Au collisions at
  $\sqrt{s_{NN}}=200$~GeV. Results are shown with three different
  freeze-out parameter values.}
\end{minipage}\hspace{2pc}%
\begin{minipage}{16pc}
\includegraphics[width=16pc]{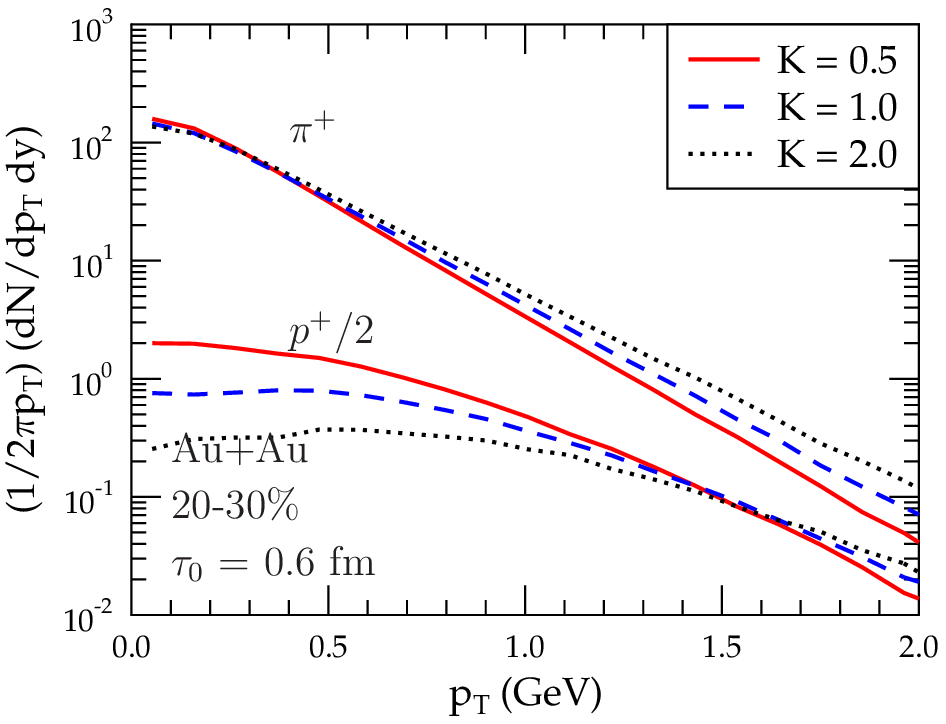}
\caption{\label{fig: chance K spectra} Spectra of positively charged
  pions and protons in 20-30\% central Au+Au collisions at
  $\sqrt{s_{NN}}=200$~GeV. Results are shown with three different
  freeze-out parameter values.}
\end{minipage}
\end{figure}

Finally, we show how the results depend on the freeze-out parameter $\Kfo$.
In Fig.~\ref{fig: chance K surface} we plot the freeze-out surfaces
with $\Kfo=0.5,1.0$ and $2.0$ and in Fig.~\ref{fig: chance K spectra}
we show the $p_T$-spectra of pions and protons with these three
values. We see that with larger value of $\Kfo$ the system lives
longer. This naturally leads to the fact that more flow is developed
and thus the spectra for pions and protons is flatter. Also because we
have chemical equilibrium until freeze-out, the number of protons
depends on the freeze-out value $K$ since the average temperature on
the surface depends on $K$. Introducing a separate chemical
freeze-out would solve this problem and this will be done in the
future.

\section{Results with fluctuating initial conditions}

In the previous section we saw that the particle spectra and elliptic
flow do not change when we switch from constant temperature freeze-out
to the dynamical one. However, we used smooth initial conditions,
where the expansion rate behaves smoothly. When we include
event-by-event density fluctuations to the initial state, the flow
develops more violently and the freeze-out criterion could affect the
final spectra.

In the first calculations we used a quite small smearing parameter
$\sigma = 0.4$~fm, which means that there is a lot of structure in the
initial energy density distribution.  Consequently the expansion rate
can vary a lot during the evolution. An example of the freeze-out
surface with dynamical condition is shown in Fig.~\ref{fig: ebye
  surface 04}. We can see that many long living small scale structures
exist: Thin rod-like ``horns'' and narrow but wide ``fins''.
Conceptually this kind of structures are difficult, since their size
is smaller than the average mean free path of particles, and thus from
microscopic point of view, they shouldn't exist.

\begin{figure}[h]
\begin{minipage}{16pc}
\includegraphics[width=16pc]{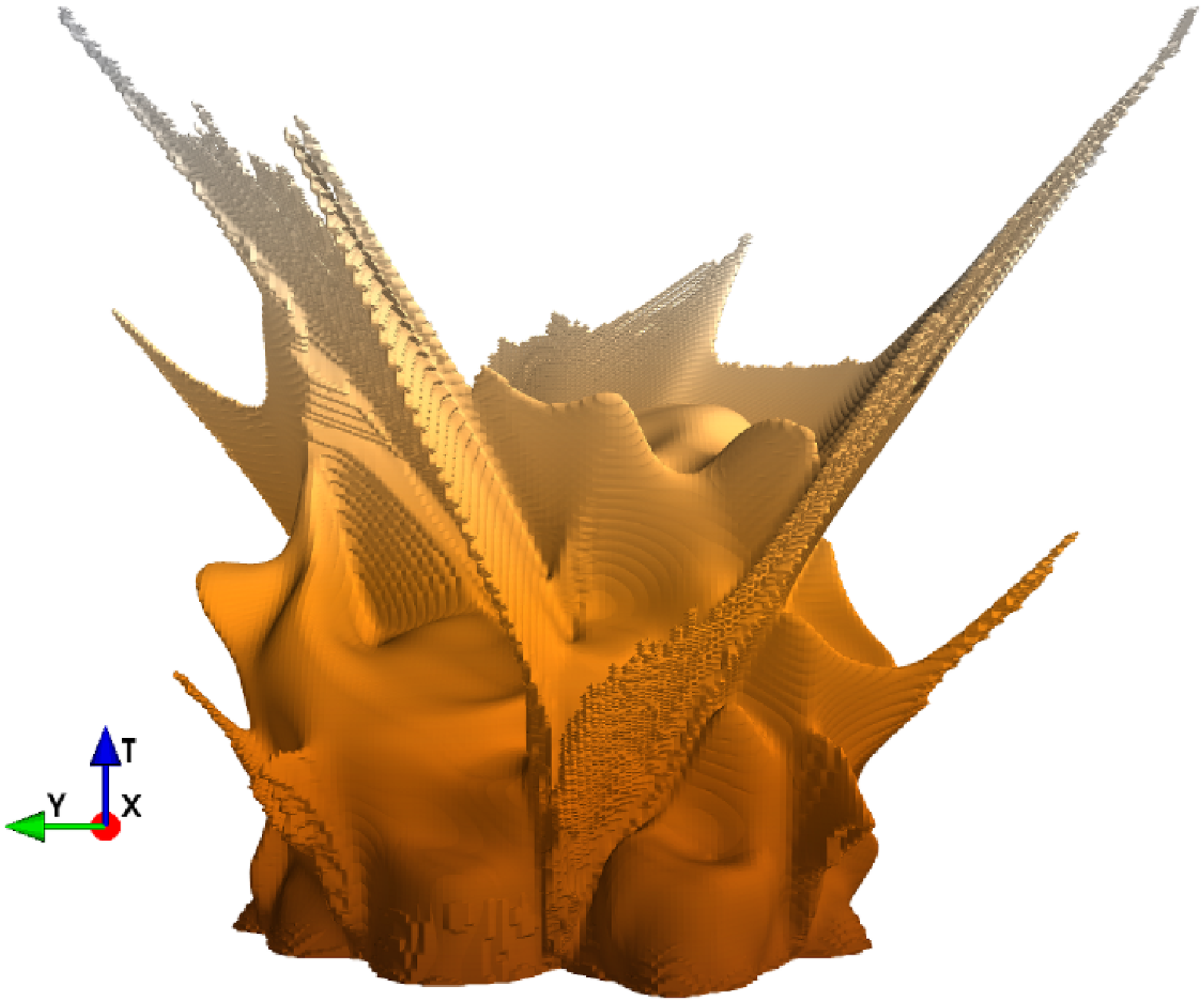}
\caption{\label{fig: ebye surface 04} Freeze-out surface with
  fluctuating initial conditions with $\sigma = 0.4$~fm. 
}
\end{minipage}
 \hspace{2pc}%
\begin{minipage}{16pc}
\includegraphics[width=16pc]{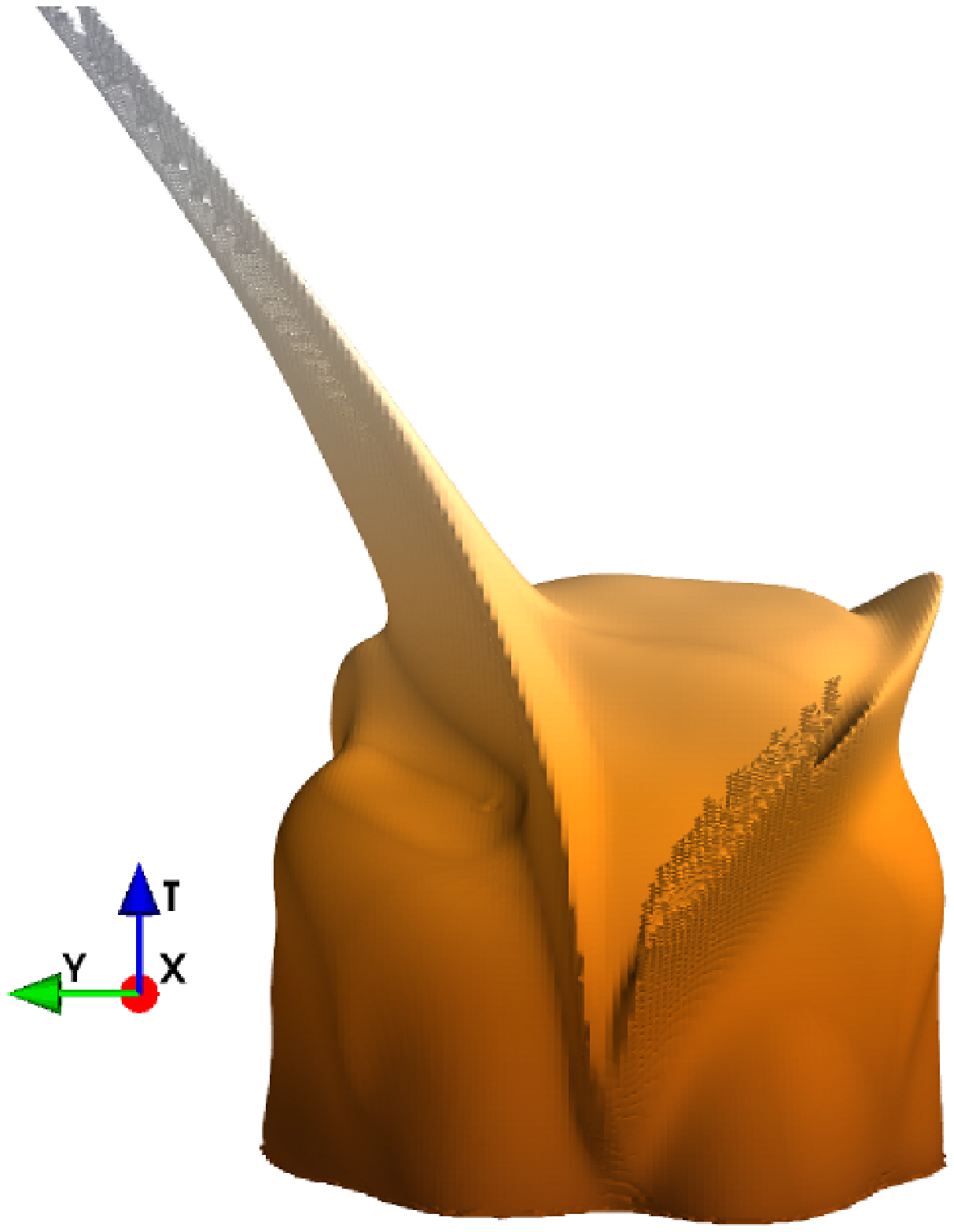}
\caption{\label{fig: ebye surface 08} Freeze-out surface with
  fluctuating initial conditions with $\sigma = 0.8$~fm. 
}
\end{minipage}
\end{figure}

The same event with double the smearing parameter is shown in
Fig.~\ref{fig: ebye surface 08}, and we immediately notice that the
surface is much smoother and only one fin remains. Thus before
studying the effects of the fins and horns, one can use smoother
initial states to check if the freeze-out condition is more important
when the fluctuating initial state is used than when an averaged
initial state is used. More advanced studies where the effect of the
fins and horns are explored are left for future work.

The reason for the fin and horn formation is that with fluctuations
there are some regions where the expansion rate goes to negative,
\emph{i.e.}\ there is compression. By definition the dynamical
freeze-out condition does not allow those regions to freeze-out, no
matter how long living or cold they are. It is obvious that the
smaller the smearing parameter $\sigma$, the more pronounced the
initial peaks and valleys are in the initial density profile, and the
more common the regions are where the fluid compresses instead of
expanding.  Thus the freeze-out surface is much more complicated in
Fig.~\ref{fig: ebye surface 04} than in Fig.~\ref{fig: ebye surface 08}.

\begin{figure}[h]
\begin{minipage}{16pc}
\includegraphics[width=16pc]{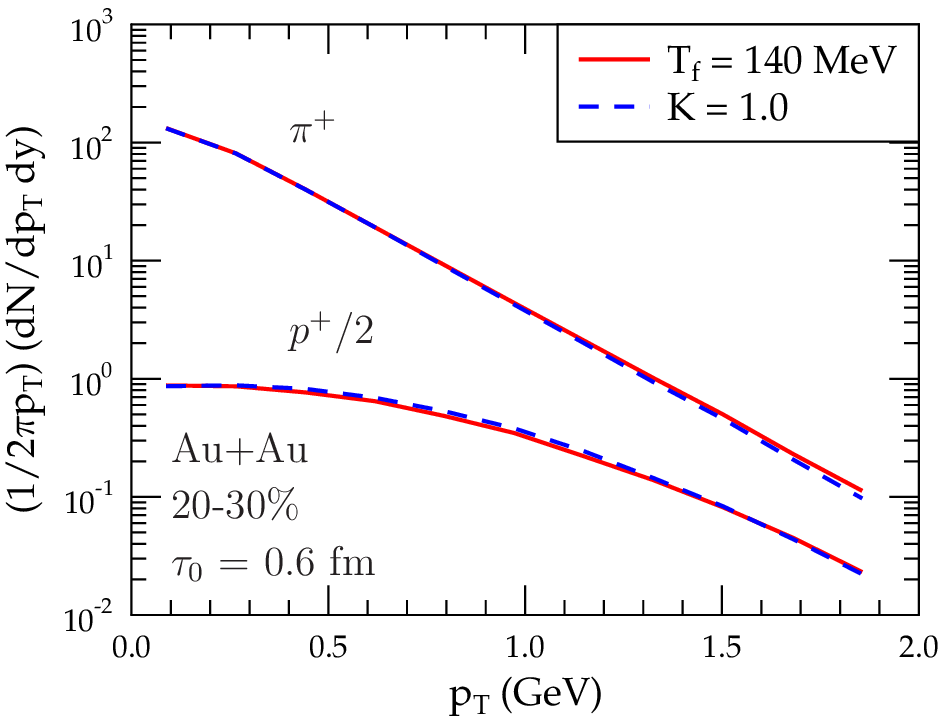}
\caption{\label{fig: spectra ebye} Spectra of positively charged pions
  and protons in 20-30\% central Au+Au collisions at
  $\sqrt{s_{NN}}=200$~GeV. Results are shown with both dynamical and
  constant temperature criterion.}
\end{minipage}
 \hspace{2pc}%
\begin{minipage}{16pc}
\includegraphics[width=16pc]{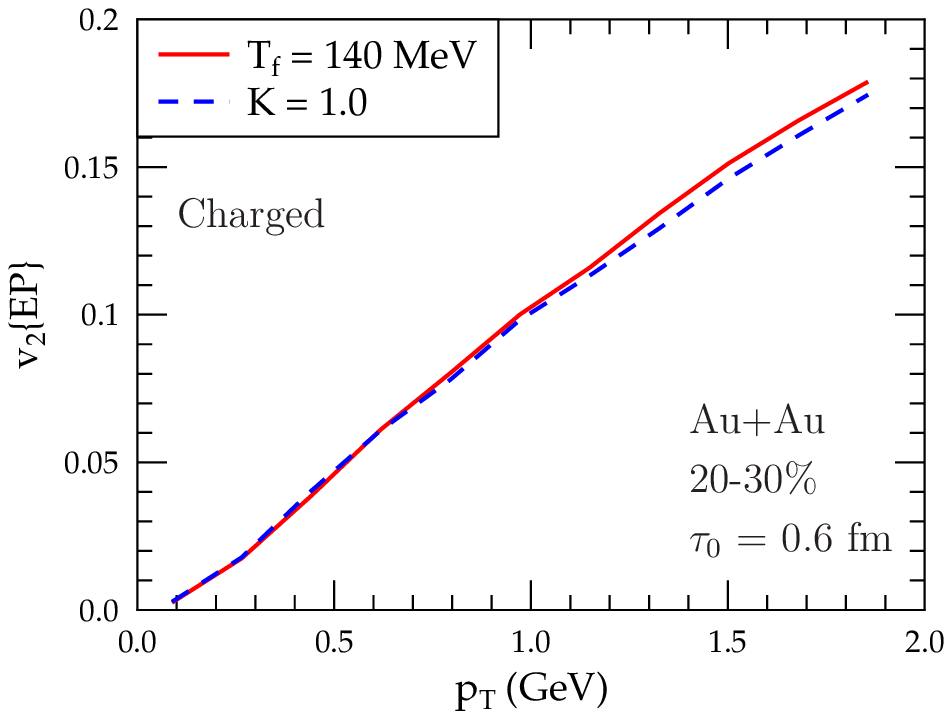}
\caption{\label{fig: v2 ebye} Elliptic flow of charged particles
  calculated with event plane method in 20-30\% central Au+Au
  collisions at $\sqrt{s_{NN}}=200$~GeV. Results are shown with both
  dynamical and constant temperature criterion.}
\end{minipage}
\end{figure}

In Fig.~\ref{fig: spectra ebye} we plot the spectra of pions and
protons and in Fig.~\ref{fig: v2 ebye} we plot the elliptic flow of
charged hadrons. Here $v_2$ is calculated with event plane method in
the same way as in Ref.~\cite{Holopainen:2010gz}. We can see that also
in this situation there are no differences between the two freeze-out
conditions. The situation is similar with and without initial state
fluctuations, although the reason for this may be that we had to use
very smooth event-by-event initial states in order to get rid off the
multiple fins and horns, which we consider unphysical.

\section{Conclusions}

We have argued that the constant temperature freeze-out is an
oversimplification, but the differences compared to the dynamical
freeze-out criterion are small in particle spectra and elliptic
flow. Freeze-out surfaces itself are different, but effects from the
increased temperature and decreased radial flow at the edges in
dynamical freeze-out cancel each other and finally there are no
visible effects in the particle spectra and elliptic flow.

However, here our event-by-event studies were limited to the case
where initial fluctuation smearing parameter was large. Because of
that our initial states are very smooth and we can expect that the
behavior is very similar to the optical Glauber model case. With
smaller smearing parameters the freeze-out surface had many small
scale structures (which we called horns and fins) which are most
probably unphysical. Thus we did not evaluate the particle
distributions yet in that case. Ultimately one needs to evaluate the
spectra in this case too, but before that we have to devise a way to
distinguish the effect of these small scale structures on the spectra
to know when the change in the spectra might be unphysical.

On the other hand it has been shown that finite viscosity smoothens
the density distribution at the end of the evolution
considerably~\cite{Schenke:2010rr}. Thus the small scale structures
may disappear altogether in a viscous calculation, and it would be
interesting and easier to apply dynamical freeze-out criterion in the
context of viscous hydrodynamics.

The work of H.H.\ was supported by the Extreme Matter Institute (EMMI)
and the work of P.H.\ by BMBF under contract no.\ 06FY9092.

\end{document}